# Gate-Tunable Semiconductor Heterojunctions from 2D/3D van der Waals Interfaces

Jinshui Miao,[1] Xiwen Liu,[1] Kiyoung Jo[1], Kang He[1], Ravindra Saxena,[1] Baokun Song,[1] Huiqin Zhang,[1] Jiale He,[2] Myung-Geun Han,[3] Weida Hu,[2] Deep Jariwala[1,*]

[1]Electrical and Systems Engineering, University of Pennsylvania, Philadelphia, PA, USA

[2]Shanghai Institute of Technical Physics, Chinese Academy of Sciences, Shanghai 200083, China

[3]Brookhaven National Laboratory, Upton, NY, USA

Corresponding author: dmj@seas.upenn.edu

**ABSTRACT**: Van der Waals (vdW) semiconductors are attractive for highly scaled devices and heterogeneous integration since they can be isolated into self-passivated, two-dimensional (2D) layers that enable superior electrostatic control. These attributes have led to numerous demonstrations of field-effect devices ranging from transistors to triodes. By exploiting the controlled, substitutional doping schemes in covalently-bonded, three-dimensional (3D) semiconductors and the passivated surfaces of 2D semiconductors, one can construct devices that can exceed performance metrics of "all-2D" vdW heterojunctions. Here, we demonstrate, 2D/3D semiconductor heterojunctions using $MoS_2$ as the prototypical 2D semiconductor laid upon Si and GaN as the 3D semiconductor layers. By tuning the Fermi levels in $MoS_2$, we demonstrate devices that concurrently exhibit over seven orders of magnitude modulation in rectification ratios and conductance. Our results further suggest that the interface quality does not necessarily affect Fermi-level tuning at the junction opening up possibilities for novel 2D/3D heterojunction device architectures.

**ToC Figure**

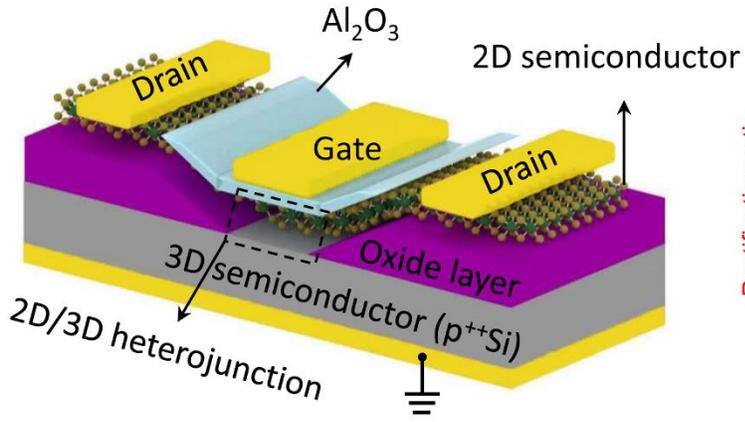
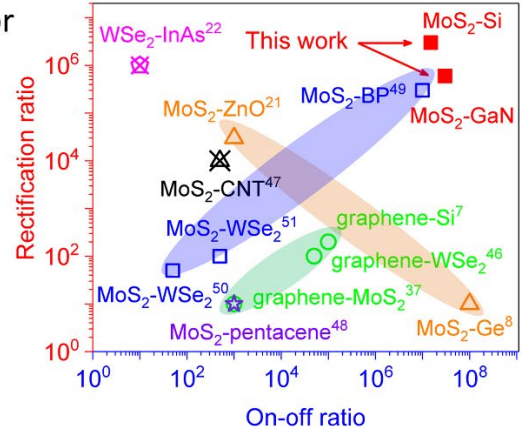

**INTRODUCTION**

The advent of van der Waals (vdW) materials has renewed enthusiasm in noveldevice designs for highly scaled field-effect devices.[1-6] This is in part due to the atomically-thin two-dimensional (2D) bodies and self-passivated surfaces that allow superior electrostatic control over all known covalently bonded three dimensional (3D) semiconductors.[7-8] This self passivation in vdW materials arises from the nature of their binding where the surface atoms in each layer have no unsaturated orbitals and hence show no tendency towards chemical bonding or reactions with the neighboring medium. As a consequence, it is significantly easy to embed a vdW 2D semiconductor between metal gates separated by thin dielectrics for strong electrostatic modulation. However, Si and III-V technology is very mature and therefore the introduction of new materials is a significant challenge. Further, vdW materials with band-gap values and material quality comparable to Si or III-V materials are not readily available.[9] In addition, stable, complementary doping in vdW semiconductors remains a persistent challenge which prevents the realization of low-power circuits.[10-15] Conversely, 3D semiconductors have well established and stable complementary doping schemes available. However, the lack of self-passivated interfaces and bulk structureprevents superior electrostatic modulation in a planar geometry.[16-17] Therefore, there exists an opportunity to combine 2D materials with 3D semiconductors to explore fundamental charge-transport phenomena at their interfaces and exploit them for devices.[18] This approach is particularly appealing since it is easy to transfer 2D vdW layers on 3D surfaces which can in turn help passivate the 3D surface due to the inert chemical nature of the 2D layers, thereby forming an electronically active interface.[7-8] Further, this also opens avenues to vertical integration of more devices on top of a fully fabricated 3D complementary metal-oxide semiconductor (CMOS) platform.[19-20] Along these lines, several approaches have been adopted for varying purposes ranging from low-power to high-current modulation switching.[7-8, 21-22] However, few efforts exist in making high-performance switching devices from 2D/3D heterojunctions. Early attempts to interface graphene with 3D semiconductors while successful, have been limited in performance due to the limited barrier height between semi-metallic graphene and the 3D semiconductor.[7, 23] Other approaches have been limited to exploiting low-power operation or tunneling phenomena and photoresponse in two-terminal devices.[24-29] Overall, the advantage of using ultrathin 2D materials and exploiting their field tunability has not been systematically investigated and exploited.

In this work, we address this challenge by fabricating three-terminal, gate-tunable diodes (triodes) comprising of mono to few layers of $MoS_2$ as the 2D semiconductors and degenerately doped Si and GaN as the 3D semiconductors. At the interface of these junctions, we demonstrate that Fermi-level is tunable in the $MoS_2$ to achieve highly tunable rectification ratios across seven orders of magnitude i.e from 0.1 to $10^6$. Concurrently, we have also demonstrated that these triodes can effectively serve the purpose of a switching device with on/off ratios exceeding $10^7$.

Our results suggest that the 2D/3D semiconductor heterojunction system is highly tunable, near-ideal and effective for electronic applications despite the lack of perfect passivation at the interface. Given the hybrid nature of the device, it can serve the function in both switching and rectifying applications. Therefore, this device opens new possibilities of using hybrid architectures such as diode transistor logic (DTL)[30-31] and other hybrid-logic concepts[32] all integrated on top of Si CMOS. With a wealth of 2D semiconductors now isolated, combined with the availability of controlled complementary doping in 3D semiconductors, in addition to epitaxial liftoff[33-34] and remote-epitaxy enabled layer transfer[35] techniques, our work opens up new opportunities in high-performance and multi-functional, heterostructure devices for vertical integration on conventional semiconductor architectures.

**RESULTS AND DISCUSSION**

We begin by fabricating devices on highly doped wafers of Silicon or Galium Nitride. Figure 1a shows the schematic of a p-n heterojunction consisting of a 2D/3D vdW heterostructure between a few-layer $MoS_2$ (n-type) and degenerately p-doped Si covered with an alumina gate dielectric and metal electrodes. Details of microfabrication processes are provided in Materials and Methods and Supporting information (SI) Figure S1, S2. A finite element simulation of this heterojunction system shows that the band diagram of the $p^{++}Si$-$MoS_2$ ($p^{++}Si$ resistivity ≤ 0.005 ohm-cm) heterojunction p-n diode (Fig. 1b) is a type II junction under equilibrium. Based on known work-functions and electron-affinities,[36-39] $p^{++}Si$ conduction band lies above the $MoS_2$ conduction band. Further, owing to the degenerately doped nature of Si, the depletion width exclusively resides within the $MoS_2$ part with the width equalling ~ 50 nm in our simulation. The alignment and widths agree well with previously reported values.[22, 40] This observation also suggests that the $MoS_2$ (~6 nm) used in our devices must be completely depleted in the entire overlapping junction region with $p^{++}Si$. However, a small depletion region of ~ 50 nm may exist within the $MoS_2$ flake at the boundary of the junction with Si. Other structural characterizations including optical and electron microscopy suggest the clear formation of heterojunction regions (Fig. 1c, d). Topography analysis by atomic force microscopy (AFM) reveals that the thickness of the $MoS_2$ layer (~ 6 nm, Fig. 1e) while the spatial map of tip amplitude (Fig. 1 e, inset) clearly shows the square depression below the gate electrode where the 2D/3D heterojunction forms. Additional optical and electron microscopy characterizations including cross-sectional composition analysis are provided in the SI (Fig. S2 and S3).

Upon structural characterization and evaluation of the electronic structure, direct current (DC) electrical transport measurements were performed under varying drain ($V_{DS}$) and gate ($V_{GS}$) biases. The $I_{DS}$-$V_{DS}$ output characteristics of $p^{++}Si$-$MoS_2$ heterojunction diodes at different $V_{GS}$ show strong modulation of current as shown in Figure 2a,b. This strong modulation shows no dependence as a function of $MoS_2$ thickness in the limit that we have investigated which is from 6-8 layers as seen in Figure 2 down to monolayers (see SI Fig. S4). In particular, the reverse saturation current (positive $V_{DS}$) is observed to be modulated from pA range to above 10 µA. This results in the device transitioning from a highly rectifying state at $V_{GS}$ = -12V (black line) to an ohmic-like state at $V_{GS}$ = 12V (orange line) demonstrating the gate-tunability of current through the p$^{++}$Si-MoS$_2$ heterojunction. Since the reverse current can be strongly modulated, the device can be operated as a digital logic switch when biased in this voltage range. The $I_{DS}$-$V_{GS}$ transfer characteristics (Fig. 2c) at different reverse drain biases show this on/off current modulation. It is also observed that gate-modulation characteristics of drain current in forward bias *versus* reverse bias are fundamentally different which is explained below with band diagrams. In contrast, the transfer and output characteristics of a control MoS$_2$ FET (see SI Fig. S5) built from the same flake show ohmic-like I-V behaviors suggesting that the 2D/3D junction dominates current output. Metalcontacts to the p$^{++}$Si substrate are alsoOhmic , as confirmed by the linear I-V relationship with small resistance (~44 Ω) (see SI Fig. S6). Our triode heterojunctions show on/off current ratios of 2×10$^7$ (@ $V_{DS}$ = 4V) and a rectification ratio of up to 3×10$^6$ (@ $V_{GS}$ = -12V). The comparison of on/off ratio versus $V_{DS}$ (blue) and rectification ratio versus $V_{GS}$ (red) can be seen in Figure 2d. The rectification ratio varies by six orders of magnitude with $V_{GS}$ while the on/off ratio varies by seven orders of magnitude with $V_{DS}$. We further note that this superior performance arises from the highly asymmetric nature of the junction that is obtained by the use of p$^{++}$ Si. This allows the depletion region to be completely concentrated in the MoS$_2$ which is effectively modulated by the gate. In addition, the lack of Fermi-level pinning due to the passivated vdW nature of the MoS$_2$ further contributes to effective gate-tunability. Our triode device can serve more functions than simply serving as a rectifying diode (for negative $V_{GS}$ values) or transistor (for reverse $V_{DS}$ values). In particular, under forward bias, there is minimal current modulation as a function of gate voltage (Figure 2c, $V_{DS}$ = -2 to -4V and Figure S7). This suggests that the triode can be used as a tunable current source whose magnitude is dictated purely on $V_{DS}$ value which is of particular importance in analog circuits and LED drivers for noise-induced fluctuations.[41]

To develop a mechanistic understanding of the device operation, we investigate the junction band-diagrams under various biasing conditions by finite-element simulations, as shown in Figure 2e, f. For $V_{GS}$ > 0V, the Fermi level in MoS$_2$ is pulled up, while the Fermi level in p$^{++}$Si is pinned below the valence band (Figure 1e (i)). As a result, a p-n junction is formed and the built-in potential barrier for electrons and holes is lowered with $V_{DS}$ < 0V, leading to a large current dominated by diffusion. For $V_{DS}$ > 0V, the conduction band of MoS$_2$ is pushed further below the valence band of p$^{++}$Si, narrowing the depletion region further to ~ 10 nm. In this state the electrons from the valence band of p$^{++}$Si can directly tunnel into the conduction band of MoS$_2$ due to the availability of density of states (DOS) in the MoS$_2$ conduction band, resulting in high current flow (Fig. 2e (ii)). When the MoS$_2$ is depleted to near intrinsic levels ($V_{GS}$ < 0V), the Fermi level moves to the middle of band-gap forming an i-p$^{++}$ junction which when forward biased ($V_{DS}$

< 0) narrows down the depletion region and produces a large diffusion dominated current (Figure 2f (i)). The converse is true under reverse bias ($V_{DS} > 0V$) which produces a wide depletion region almost exclusively in the $MoS_2$ which not only increases the width for direct tunneling but also reduces availability of electron states in $MoS_2$ leading to a low off-state current (Fig. 2f (ii)).

While the above results of I-V characteristics suggest superior rectification and modulation, the fundamental nature of charge injection at a 2D/3D interface remains to be determined. To ascertain the current flow in $p^{++}Si$-$MoS_2$ heterojunction diode, we first characterize the current as a function of the metal contact length and 2D/3D junction area. The fabrication process is detailed in SI; Figure S8 and S9. $I_{DS}$-$V_{DS}$ output characteristics with varying contact lengths (Figure 3a) show that the current decreases by approximately ∼ 700% as the contact length changes from 18 μm to 2 μm. On the other hand, the output characteristics *versus* 2D/3D junction area (Figure 3c) show that current decreases slowly by only ∼ 200% as the junction area changes from 169 μm$^2$ to 36 μm$^2$. Device conductance *versus* contact length at $V_{DS}$ = -4V extracted from Figure 3a is shown in Figure 3b which further demonstrates that the conductance linearly depends on contact length with a slope of ∼4μS/μm. However, the perimeter-normalized conductance is independent of the junction area (slope of ∼0.1μS/μm$^2$) and it shows no change for varying junction areas as shown in Figure 3d. Based on these electrical characteristics we conclude that although the 2D/3D interface is planar in nature, the current injection is primarily one dimensional (1D) because of lateral drain contact geometry. To confirm the 2D/3D nature of the vdW junction, we fabricated and measured two-terminal vertical transport devices with varying areas of top drain electrodes (Figure 3e (inset)). In these devices, the conductance changes by 184.4% as the drain electrode area changes by 178.6% suggesting that the device conductance linearly depends on vertical 2D/3D junction area as shown in Figure 3e and f. Further, the area-normalized conductance stays constant at ∼10 μS/μm$^2$ (Figure 3f) confirming the uniformity and electronic homogeneity of the 2D/3D contact. We have further verified our claims of electronic homogeneity and uniformity of the 2D $MoS_2$ /3D $p^{++}$ Si contact by measuring multiple vertical-junction devices of constant contact area (SI, Figure S10) and performing spatial mapping of conductance using conductive AFM measurements (SI, Figure S11) both of which agree with the above-discussed results and solidify our claim.

To get more direct evidence and further reinforce our observations from electrical measurements, we perform photocurrent measurements to deduce the electronically active area of the junction. Photocurrent microscopy shows a large photocurrent near the edge of 2D/3D junction (Figure 3g (ii)) because $MoS_2$ collects photo-generated electrons and Si collects photo-generated holes leading to the photocurrent (see SI Figure S12). . A photocurrent signal suggests that the illuminated junction spot is electronically active as excited electrons and holes can separate, drift out and get collected at the electrodes under zero bias. By reciprocity, the same regions would participate in diffusive current transport under forward and reverse bias conditions in the absence of illumination. This observed photocurrent signal diminishes further away from the 1D edge created by the etched oxide at which the 2D/3D junction initiates. The photocurrent while measurable along the edges up until the points where the metal electrodes extend along the length of the square. The photocurrent is however, more than three times lower at the center of the square (Figure 3g(ii) and (iii)) suggesting negligible carrier separation and collection from the center. This observation provides a direct illustration that the carrier separation/collection and hence injection in a 2D/3D junction primarily occurs at a 1D interface as confirmed by 1D square shape in Figure 3g (ii). It is also worth noting that no measurable photocurrent is observed from the nonoverlapping region of $MoS_2$ or the metal contact, suggesting that there is minimal contribution from drift or metal/$MoS_2$ contact induced charge separation, further indicating that the current flow in the $p^{++}Si$-$MoS_2$ junction is dominated by the junction (Figure 3g (iii)). Therefore, we conclude that the device output current is not only dominated by the junction but more importantly the carrier injection occurs predominantly at the 1D edge of 2D/3D junction. This observation concurs related observations in the electronic transport studied of metal contacts to 2D materials. 3D metal contacts to 2D TMDCs and graphene are known to have a 1D edge length dependence.[42] However, often in the case of 3D metal contacts, there is covalent bonding involved.[43] In contrast, this is the first demonstration of this kind for a 2D/3D vdW semiconductor junction wherein we observe that even in a vdW 2D/3D contact, the carrier injection is mainly 1D. This observation also has further implications in terms of areal scaling of the device. As long the device geometry entails charge injection and collection via metal side contacts, the 2D/3D junction area shouldn't matter and thus arbitrarily narrow 2D/3D junction devices can be packed in a given area to achieve large packing density or current density.

Finally, to establish comparative performance in a 2D/3D interface, we also use 3D GaN to build gate-tunable 2D/3D heterojunction diodes (Figure 4a and Figure S13, S14 in SI). The motivation to use GaN is two-fold: a. does isotype ($n^{++}/n$) versus anisotype ($p^{++}/n$) heterojunctions have an impact on a 2D/3D junction performance? and b. does the interface atomic and electronic structure affect field modulation characteristics? Figure 4b shows the gate-dependent output current ($I_{DS}$) as a function of applied drain bias ($V_{DS}$) on $MoS_2$ while the contact to GaN is grounded. Similar to the silicon case, the GaN-based device exhibits a high rectification ratio of up to ~ $6 \times 10^5$ at $V_{GS}$ = -12V and it varies by over six orders of magnitude as a function of gate voltage. Figure 4c shows the drain current ($I_{DS}$) as a function of gate voltage ($V_{GS}$) with drain bias changing from -4 to 4V. It is observed that the device on/off ratio is higher at positive drain biases ($V_{DS}$ = 1 to 4V) than that at negative drain biases ($V_{DS}$ = - 1to -4V). It is worth nothing that the switching mechanism is somewhat different for the GaN/$MoS_2$ isotype ($n^{++}/n$) junctions in comparison to Si/$MoS_2$ anisotype ($p^{++}/n$) junctions as detailed below with simulated band diagrams. Further, we observe the ideality factor of GaN-$MoS_2$ diodes (1.5 < n < 2.0) is smaller than that of Si-$MoS_2$ diodes (2.0 < n < 3.5) because, in contrast to GaN, Si surface has a thin native oxide layer which degrades interface quality of 2D/3D junction (Figure 4d, Figure S3, Figure S15 in SI).[26, 44-45] Despite the differences in quality of the interface, the gate-tunability of rectification ratio maintains the same trend (Figure 4e, red plot) and the rectification ratio reaches a maximum when the MoS$_2$ is fully depleted suggesting a maximum in built-in potential and formation an n$^{++}$/i junction. Likewise, a high asymmetry is observed in gate-dependent conductance (transfer characteristics) depending on whether the diode is forward biased (-V$_{DS}$) versus reverse biased (+V$_{DS}$). This is once again a distinguishing feature of this device that the drain current remains constant as a function of V$_{GS}$ under forward bias. Even under reverse bias, at high positive values of V$_{GS}$, the drain current reaches saturation (Figure 4c). We believe this is the case since one side of the junction is fully depleted. In summary, a high-performance three-terminal device with a record-high on/off ratio and diode with a high rectification ratio is simultaneously achieved in a single GaN/MoS$_2$ heterostructure, as summarized in Figure 4e. It is worth noting however that several attempts have been made at making gate-tunable vdW heterojunctions.[7-8, 21-22, 37, 46-51] In most 2D/2D or 2D/3D cases, the doping levels in both semiconductors are simultaneously moving due to the semitransparency of 2D layer to electric fields. Therefore either these devices were limited by off-currents or on-currents (hence a trade-off in their ratio) or built-in potential (rectification ratio). If we fix the doping level in one semiconductor and make it 3D, a maximum in both can be concurrently achieved in the same device which is the case here. This is shown as a comparison of the rectification ratio and on/off ratio with values reported in the literature (Fig. 4f).

Simulated band-diagrams under gate-induced accumulation (Figure 4g) and depletion (Figure 4h) of MoS$_2$ further elucidate the operation of this triode device. Unlike the case of p$^{++}$Si-MoS$_2$, simulated band alignments of n$^{++}$-GaN and n-MoS$_2$ show their Fermi levels and built-in potential barriers inversely aligned due to the comparable electron affinities of MoS$_2$ and GaN.[45] This also suggests that n-doped MoS$_2$ can serve as an efficient electron injection or collection contact for GaN devices. Under positive gate voltage (V$_{GS}$ > 0V), the MoS$_2$ is under accumulation and the Fermi level is raised close to its conduction band and the Fermi level in n$^{++}$-GaN is equal to or above its conduction band. There is no potential barrier for electrons in the MoS$_2$ conduction band which can diffuse unhindered into the GaN conduction band at V$_{DS}$ < 0V, resulting in a high current (Fig. 4g (i)). Even for reverse bias (V$_{DS}$ > 0), there is no barrier for electrons in this junction which leads to a high drift current for electrons (Fig. 4g (ii)). In fact, for sufficiently high V$_{DS}$ (> 2V), the current in reverse bias (drift) exceeds the current in forward bias without causing a breakdown and conceptually inverting the diode rectification (Fig. 4b, c). Under depletion of MoS$_2$ i.e. V$_{GS}$ < 0V, the Fermi level in MoS$_2$ shifts to the middle of band-gap while the Fermi level in n$^{++}$GaN is still pinned above the conduction band owing to the lack of electric field modulation in a heavily doped 3D semiconductor with a high DOS. For V$_{DS}$ < 0 (forward bias) once again there is a large diffusion current due to lowering of the junction barrier and there is a small reduction in magnitude due to depletion of the MoS$_2$ (Figure 4b, black plot). However, this diffusion current will have a greater contribution from holes in MoS$_2$ since the Fermi level is in the center of the gap (Figure 4h (i)) as opposed to an electron dominated current in the previous case of MoS$_2$ under accumulation (V$_{GS}$ > 0). Finally, the MoS$_2$ part is fully depleted at V$_{GS}$ < 0V and V$_{DS}$ > 0V (reverse-biased diode), which results in a very small drift current (Figure 4h (ii)).

**CONCLUSIONS**

In conclusion, we have demonstrated gate-tunable heterojunction diodes through vdW integration of 2D and 3D semiconductors. Owing to the ultrathin nature of 2D semiconductors and fixed doping of the 3D part, charge-carriers across the 2D/3D junction can be effectively modulated by a capacitively coupled gate, enabling wide tunability of diode rectification ratio of up to $10^6$ concurrently with an on/off current ratio > $10^7$ enabling utility for tunable rectification as well as digital switch. Also, we find that charge injection at a 2D/3D junction occurs along with a 1D contact interface with implications on scalability. Finally, the wide tunability of the junction and small ideality factors (1.5<n<3.0), despite an unpassivated 3D semiconductor surface, suggests a path towards heterogeneous integration on top of Si CMOS at low temperatures. In particular, with the availability of remote epitaxy and epitaxial lift-off techniques to lift and peel Si and III-V 3D semiconductors, the possibilities for integrating 2D layers with Si or III-V on arbitrary substrates towards multifunctional sensors, memory or optoelectronic layers in conjunction with a high-performance computing part poses to be a promising direction for the field in years to come.

**Acknowledgments**

D.J. acknowledges primary support for this work by the Penn Engineering Start-up funds (J.M., X.L.) and U.S. Army Research Office under contract number W911NF-19-1-0109 (K.J. and H.Z.). D.J. also acknowledges support from the NSF supported University of Pennsylvania Materials Research Science and Engineering Center (MRSEC) (DMR-1720530) and National Science Foundation (DMR-1905853). K.H. acknowledges Masters Scholar Award from the Department of Materials Science and Engineering at Penn. This work was carried out in part at the Singh Center for Nanotechnology at the University of Pennsylvania which is supported by the National Science Foundation (NSF) National Nanotechnology Coordinated Infrastructure Program grant NNCI-1542153. The work at Brookhaven National Laboratory was supported by the U.S. Department of Energy, Office of Science, and Office of Basic Energy Sciences, under Contract No. DE-SC0012704. FIB sample preparation was performed at the Center for Functional Nanomaterials, Brookhaven National Laboratory.

**Notes**

The authors declare no competing financial interests.

**Supporting Information**

The Supporting Information is available free of charge on the ACS Publications website which includes experimental methods, additional experimental data, calculations, and analysis.

**Figure 1**

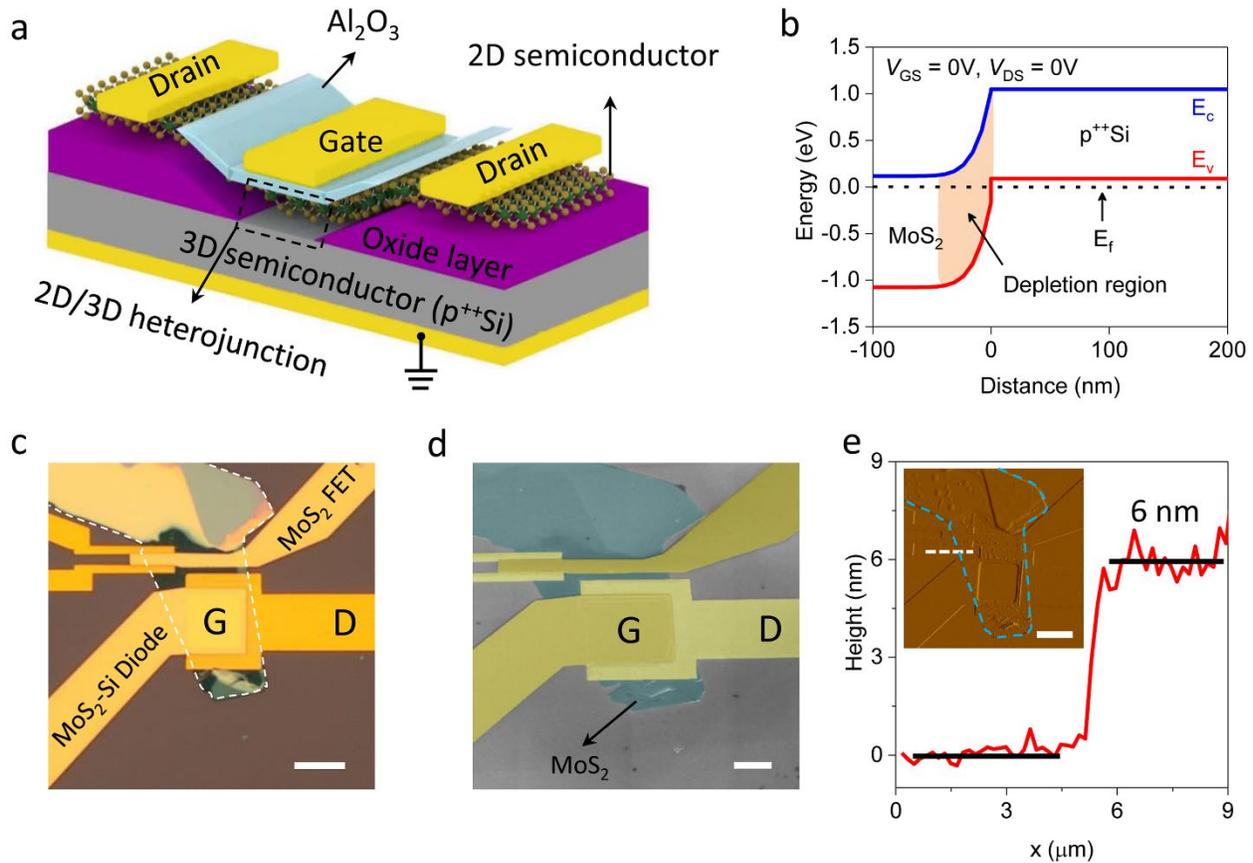

**Figure 1. 2D/3D heterojunction diode and characterization**. (a) Schematic illustration of a gate-tunable diode based on 2D MoS$_2$ and 3D Si van der Waals heterostructure. The 2D layer drapes over the etched oxide recess to form direct contact with the 3D semiconductor wafer as indicated with the dashed line box. Source and drain electrodes comprise of Ti/Au (10nm/40nm). The gate oxide is 30 nm of Al$_2$O$_3$ via atomic layer deposition and the gate electrode is also Ti/Au (10nm/40nm). (b) Simulated band diagrams of p$^{++}$Si-MoS$_2$ heterojunction diodes under equilibrium showing conduction band (E$_c$), valence band (E$_v$), Fermi level (E$_f$) and depletion region. Si bandgap (E$_g$): 1.1 eV, MoS$_2$ bandgap (E$_g$): 1.2 eV, Si electron affinity ($\chi$): 4.01 eV, MoS$_2$ electron affinity ($\chi$): 4.1 eV. (c) Optical micrograph of a representative p$^{++}$Si-MoS$_2$ heterojunction diode. MoS$_2$ flake is indicated by a white dashed line. Scale bar, 10 µm. (d) False-colored SEM image of the device. MoS$_2$ flake and electrodes are indicated by light blue and yellow color, respectively. Gate (G) and Drain (D) electrodes are appropriately labeled while the source electrode is the p$^{++}$Si substrate. Scale bar, 5 µm. (e) AFM height profile across the MoS$_2$ flake as measured from a topography map. The inset provides the device AFM tip amplitude image concurrently acquired with the topography image with the corresponding height profile region indicated by the white dashed line. The MoS$_2$ flake is indicated by a blue outline. Scale bar, 10 µm.

**Figure 2**

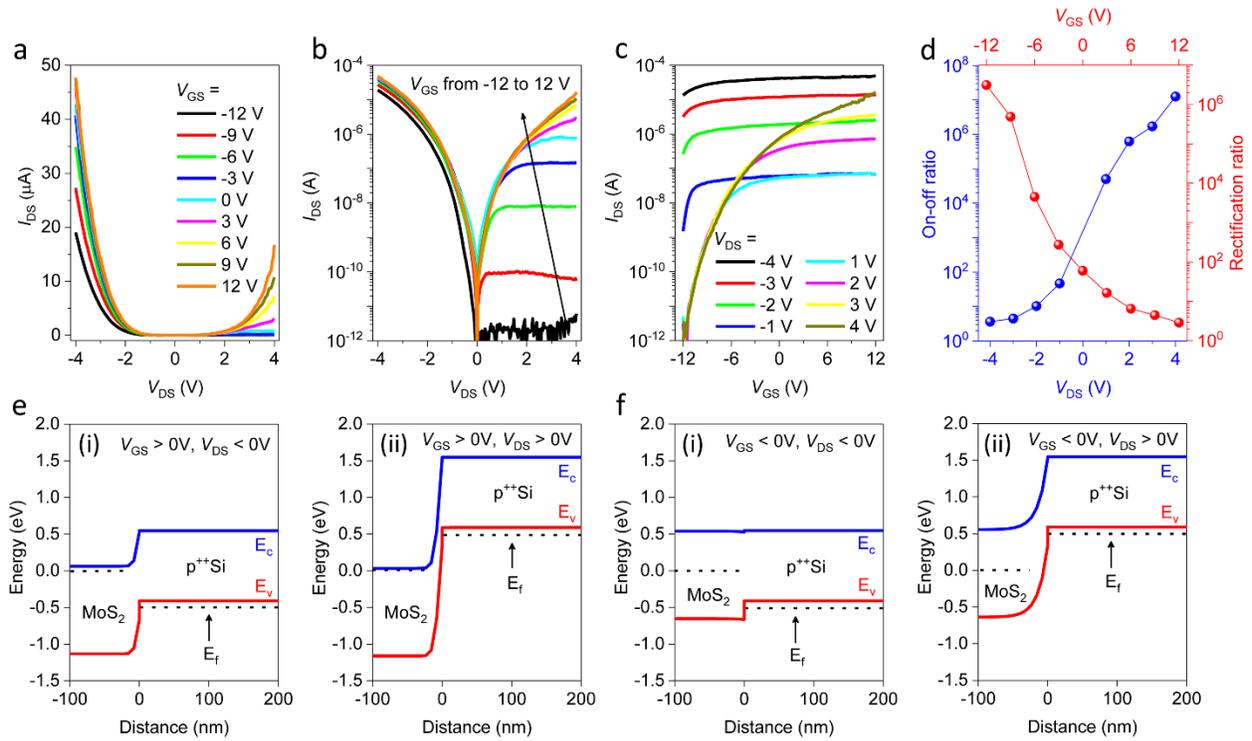

**Figure 2. Room-temperature electrical characterization of p$^{++}$Si-MoS$_2$ heterojunction diodes.** (a) Linear scale output characteristics of the device at various gate voltages showing a clear transition from a highly diode-like rectifying behavior at $V_{GS}$ = -12V to an almost symmetric $I_{DS}$-$V_{DS}$ behavior at $V_{GS}$ = 12V. The drain bias is on MoS$_2$ and the contact to p$^{++}$Si (source) is grounded. (b) Semilogarithmic scale output characteristics of the device. $V_{GS}$ varies in the range of -12 to 12V, with a step size of 3V. The highly rectifying behaviors of the I-V characteristic in addition to exponential modulation of reverse saturation current with $V_{GS}$ can be seen. (c) Transfer characteristics of the device at various drain biases. The characteristics show a clear n-type transistor behavior. The high on/off ratio at $V_{DS}$ > 0V and negligible modulation at $V_{DS}$ < 0V can be seen. (d) The device on/off ratio versus $V_{DS}$ (blue) and rectification ratio versus $V_{GS}$ (red). (e) Simulated band diagrams of p$^{++}$Si-MoS$_2$ heterojunction for (i) $V_{GS}$ > 0V and $V_{DS}$ < 0V and (ii) $V_{GS}$ > 0V and $V_{DS}$ > 0V. (f) Simulated band diagrams of p$^{++}$Si-MoS$_2$ heterojunction for (i) $V_{GS}$ < 0V and $V_{DS}$ < 0V and (ii) $V_{GS}$ < 0V and $V_{DS}$ > 0V.

**Figure 3**

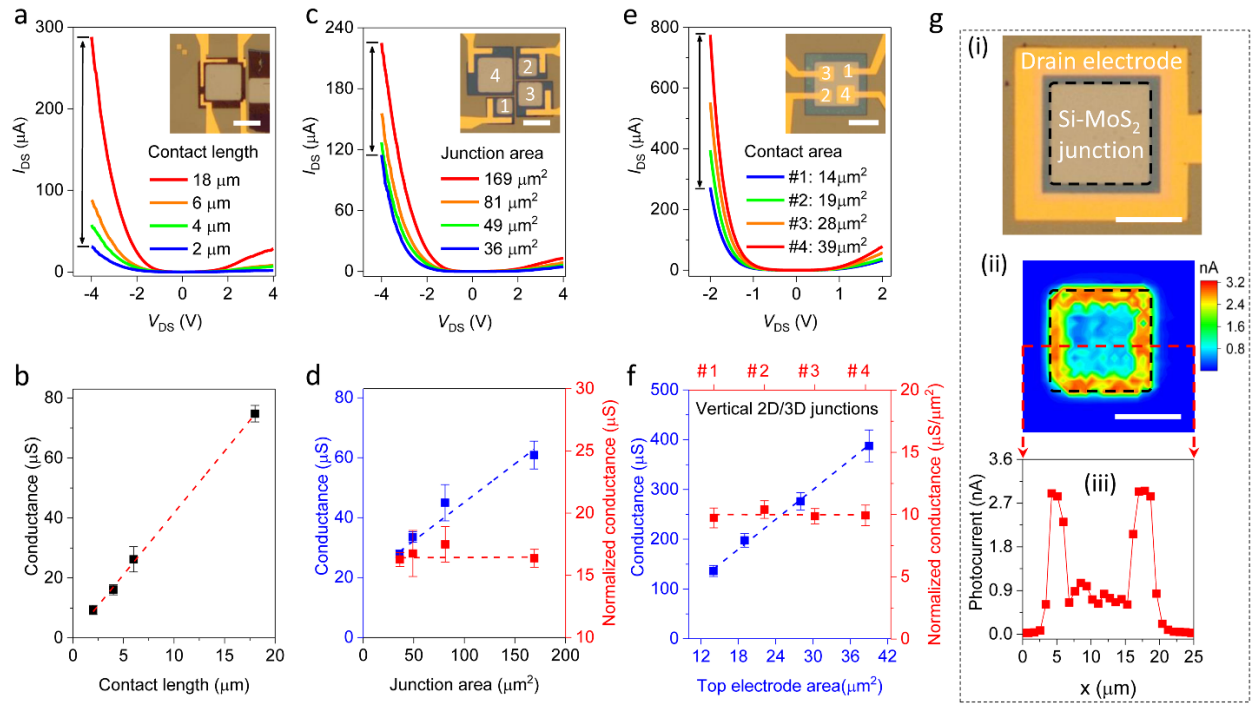

**Figure 3. Lateral and Areal Scaling in p$^{++}$Si-MoS$_2$ heterojunction diode.** (a) Output characteristics of the device with different metal contact lengths (L = 2, 4, 6, and 18 µm). Inset: optical micrograph of the device. The light grey square represents the overlapping p$^{++}$Si-MoS$_2$ junction. Scale bar, 5 µm. (b) Device conductance (@ $V_{DS}$ = -4 V) as a function of metal contact length extracted from Fig. 3a. The error bars result from the averaging of seven measurements. (c) Output characteristics of the device with different junction areas (A = 36, 49, 81, and 169 µm$^2$). Inset: optical micrograph of four p$^{++}$Si-MoS$_2$ heterojunction devices which are marked as 1, 2, 3 and 4. The light grey squares represent the overlapping p$^{++}$Si-MoS$_2$ heterojunction regions. Scale bar, 5 µm. (d) Conductance (blue) and perimeter-normalized conductance (red) as a function of junction area extracted from Fig. 3c. The error bars represent standard deviations from seven such measurements. The perimeters of four 2D/3D junctions (inset, Fig. 3c) are 24, 28, 36 and 52 µm. And the contact electrode length of the four 2D/3D junctions (inset, Fig. 3c) is 14 µm. Perimeter-normalized conductance = (conductance ($I_{DS}/V_{DS}$) × contact electrode length)/(perimeter of 2D/3D junction). (e) Output characteristics of the device with various drain electrodes (A = 14, 19, 28, and 39 µm$^2$) fabricated onto the same 2D/3D heterostructure. Inset: optical micrograph of the device with four drain electrodes which are marked as 1, 2, 3 and 4. The light grey square represents the overlapping p$^{++}$Si-MoS$_2$ heterojunction region. Scale bar, 5 µm. (f) Conductance (blue, @ $V_{DS}$ = -2 V) and area-normalized conductance (red) extracted from Fig. 3e. The error bars represent standard deviations from four such measurements. (g) Optical

micrograph (i), photocurrent map (ii) and photocurrent *versus* laser position (iii) of a representative 2D/3D junction device for 633nm laser excitation. Scale bar, 10 µm.

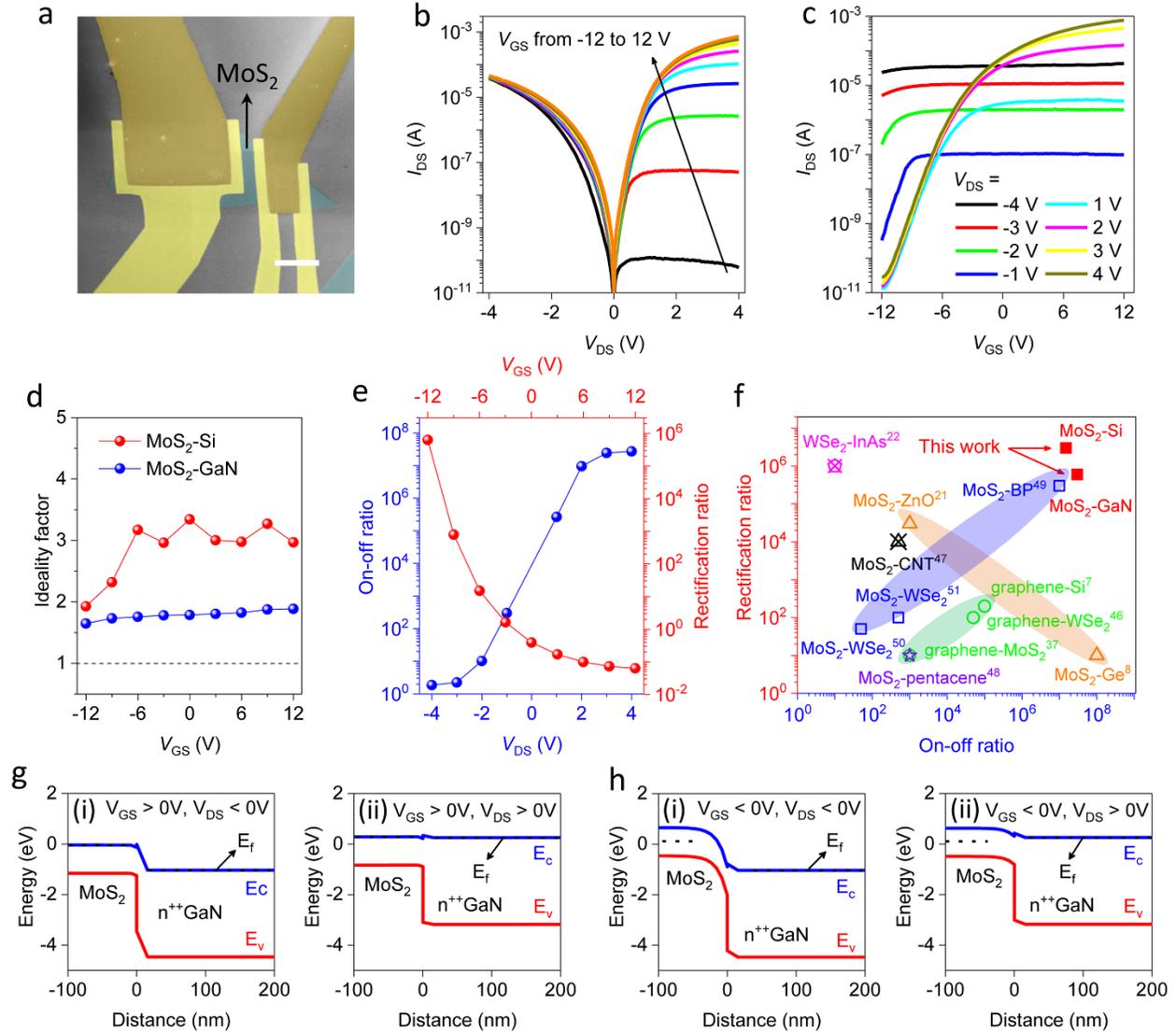

**Figure 4. Room-temperature electrical characterization of n$^{++}$GaN-MoS$_2$ heterojunction diode.** (a) False-colored SEM image of a representative n$^{++}$GaN-MoS$_2$ heterojunction diode. MoS$_2$ flake and electrodes are indicated by light blue and yellow color, respectively. Scale bar, 10 μm. (b) Output characteristics of the device at various gate voltages showing a clear transition from a highly rectifying state at $V_{GS}$ = -12V to an almost symmetric $I_{DS}$-$V_{DS}$ behavior at $V_{GS}$ = 12V. $V_{GS}$ varies in the range of -12 to 12V, with a step size of 3V. (c) Transfer characteristics of the device at various drain biases showing a clear n-type transistor behavior. The device exhibits an ultrahigh gate modulation under positive drain bias ($V_{DS}$ > 0V). In particular, under forward bias ($V_{DS}$ = -2 to -4V), there is little or no current modulation as a function of gate voltage. (d) Ideality factors of n$^{++}$GaN-MoS$_2$ (blue) and p$^{++}$Si-MoS$_2$ (red) diodes as a function of gate voltage. (e) The device on/off ratio versus $V_{DS}$ (blue) and rectification ratio versus $V_{GS}$ (red). (f) Comparison of on/off

ratio and rectification ratio with values reported in the literature. (g) Simulated band diagrams of n$^{++}$GaN-MoS$_2$ heterojunction for V$_{GS}$ > 0V, V$_{DS}$ < 0V (i) and V$_{GS}$ > 0V, V$_{DS}$ > 0V (ii), where the applied drain bias is on MoS$_2$ and the contact to GaN is grounded. (H) Simulated band diagrams of n$^{++}$GaN-MoS$_2$ heterojunction for V$_{GS}$ < 0V, V$_{DS}$ < 0V (i) and V$_{GS}$ < 0V, V$_{DS}$ > 0V (ii), where the applied drain bias is on MoS$_2$ and the contact to GaN is grounded. GaN bandgap (E$_g$): 3.4 eV, GaN electron affinity ($\chi$): 4.1 eV.

Supporting Information for

# Gate-Tunable Semiconductor Heterojunctions from 2D/3D van der Waals Interfaces

Jinshui Miao,[1] Xiwen Liu,[1] Kiyoung Jo[1], Kang He[1], Ravindra Saxena,[1] Baokun Song,[1] Huiqin Zhang,[1] Jiale He,[2] Myung-Geun Han,[3] Weida Hu,[2] Deep Jariwala[1,*]

[1]Electrical and Systems Engineering, University of Pennsylvania, Philadelphia, PA, USA

[2]Shanghai Institute of Technical Physics, Chinese Academy of Sciences, Shanghai 200083, China

[3]Brookhaven National Laboratory, Upton, NY, USA

Corresponding author: dmj@seas.upenn.edu

**MATERIALS AND METHODS**

**Device fabrication**

We start by etching square grooves/windows in the thermally grown $SiO_2$ layer on the $SiO_2/p^{++}Si$ wafer ($SiO_2$ thickness: 50 nm, $p^{++}Si$ resistivity ≤ 0.005 ohm-cm). First, PMMA e-beam resist (MicroChem 495 A8) was spin-coated (4000 rpm for 80 seconds) onto the $SiO_2/p^{++}Si$ wafer followed by baking on a hot plate at 180 degrees Celsius for 10 min. Next, we used e-beam lithography to define several square patterns (5×5 µm², 10×10 µm², 15×15 µm², 20×20 µm²) in the PMMA layer followed by development using MicroChem's developer (MIBK:IPA (1:3)). Third, buffered oxide etcher ($NH_3F:HF$ (6:1)) was used to etch away oxide layer inside the PMMA square windows and bottom $p^{++}Si$ semiconductor is exposed (etching time: 1 min). Finally, the rest of PMMA resist was dissolved by warm acetone and then several square grooves were formed in $SiO_2$ layer of $SiO_2/p^{++}Si$ wafer. The as-prepared substrates were quickly transferred to the glove box ($O_2$<0.5 ppm, $H_2O$<0.5 ppm) to aviod oxidation of Si inside the square grooves. Similarly, we also etched several square grooves in $Al_2O_3$ layer in $Al_2O_3/n^{++}GaN$ wafer. Here, the $Al_2O_3$ oxide layer (thickness = 60 nm) was grown on $n^{++}GaN$ wafer by atomic layer deposition (ALD, precursor: trimethylaluminum (TMA) and water; deposition temperature: 150 degrees Celsius; 15 msec pulses of TMA and water seprated by 20 secs).

Few-layer $MoS_2$ flakes were mechanically exfoliated from bulk $MoS_2$ crystals (HQ graphene) using Scotch tape and then transferred onto PDMS stamp. Optical microscope was used to locate uniform and thin (5 to 15 nm) $MoS_2$ flakes on the PDMS stamp. $MoS_2$ flakes were then dry-transferred onto square grooves of $SiO_2/p^{++}Si$ or $Al_2O_3/n^{++}GaN$ substrate using PDMS stamp as transfer medium. All the transfer processes are done inside the glove box. Electrical contacts were patterned onto $MoS_2$ flakes using e-beam lithography of bilayer PMMA resist (MicroChem 495 A4 and A8), e-beam deposition of Ti/Au film (10nm/40nm) and lift-off processes. ALD was then used to deposit 30nm-thick $Al_2O_3$ on $MoS_2$ as top-gate dielectric. The ALD recipe/growth conditions are same as detailed above. Before ALD, 1.5nm-thick aluminium (Al) was thermally evaporated onto $MoS_2$ as ALD seeding layer. The 1.5nm-thick Al film will be naturally oxidized when deposition system is vented. Finally, top-gate electrode (Ti/Au: 10nm/40nm) was fabricated onto $p^{++}Si-MoS_2$ or $n^{++}GaN-MoS_2$ junction region using another step of lithography and evaporation.

**Band diagram simulations**

Quasi-stationary 2D TCAD Sentaurus device simulation was performed to model the band diagrams of the gate-tunable $MoS_2/Si$ and $MoS_2/GaN$ heterojunctions, where the device structures consist of a multilayer of $MoS_2$, and an $Al_2O_3$ gate oxide thickness of 30 nm (effective oxide thickness (EOT) = 14 nm). The multilayer $MoS_2$ was treated as an ultrathin 3D semiconductor for the simulation purposes, with the following material parameters used: $E_g$ =

1.2 eV, $\varepsilon = 10\ \varepsilon_0$, $\mu_n = \mu_p = 100\ cm^2/V\cdot s$, $\tau_{n,\ SRH} = \tau_{p,\ SRH} = 1$ ns, $m_e^* = m_h^* = 0.45\ m_0$, and conduction band offset of 0.1 eV and 0.2 eV were used in the $MoS_2/Si$ and $MoS_2/GaN$ heterojunctions, respectively. Poisson equation, electron continuity and hole continuity equations are primarily solved in the device simulations, where the boundary conditions are set by the source, drain, and gate biases.

**Device characterization**

Electrical measurements were performed under vacuum ($10^{-3}$ mbar) in Lakeshore vacuum probe station using Keithley 4200A semiconductor characterization system. Raman and photocurrent map were carried out using Horiba LabRAM HR Evolution instrument with a 633-nm laser and 50x long working distance objective. AFM characterizaitons were performed using an AIST-NT SPM SmartSPM$^{TM}$-1000 which was done in the tapping mode with 200 kHz resonance frequency. SEM imaging was performed using a JEOL 7500F high resolution scanning electron microscope (HRSEM) and a 5 kV accelerating voltage was used for imaging the devices. TEM sample was prepared by focused ion beam with 5 keV $Ga^+$ ions for final milling. For EDX mapping, a Talos F200X equipped with a four-quadrant 0.9-sr energy-dispersive X-ray spectrometer operated at 200 keV was used. Detailed band structure simulation was performed using Synopsys Sentaurus TCAD.

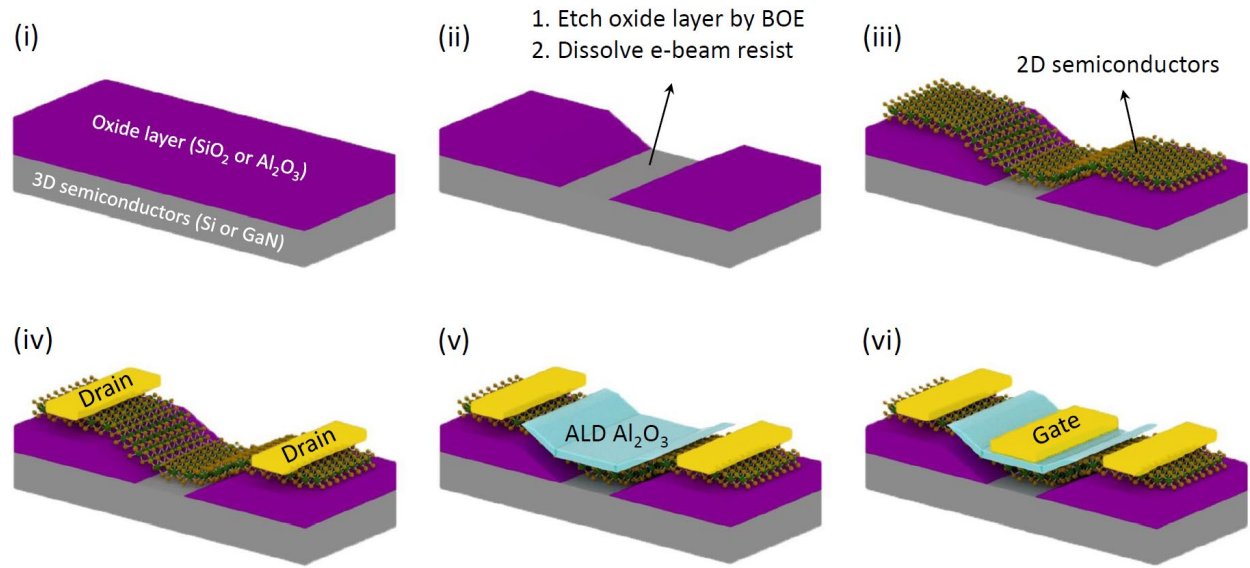

**Figure S1. Device fabrication process**. (i) Degenerately doped 3D bulk semiconductor (p$^{++}$Si or n$^{++}$GaN) with thin oxide layer on top. (ii) Square grooves were defined by e-beam lithography of PMMA and the inside oxide layer was etched away by buffered oxide etcher. (iii) Dry-transfer few-layer 2D chalcogenide semiconductors on the top of exposed 3D semiconductor. (iv) Drain electrodes (Ti/Au: 10nm/40nm) were fabricated onto 2D flake by e-beam lithography, e-beam deposition and lift-off processes. (v) Atomic layer deposition of 30-nm thick Al$_2$O$_3$ as top-gate dielectric. (vi) Top-gate electrode (Ti/Au: 10nm/40nm) was fabricated on ALD Al$_2$O$_3$ oxide layer by e-beam lithography, e-beam deposition and lift-off processes.

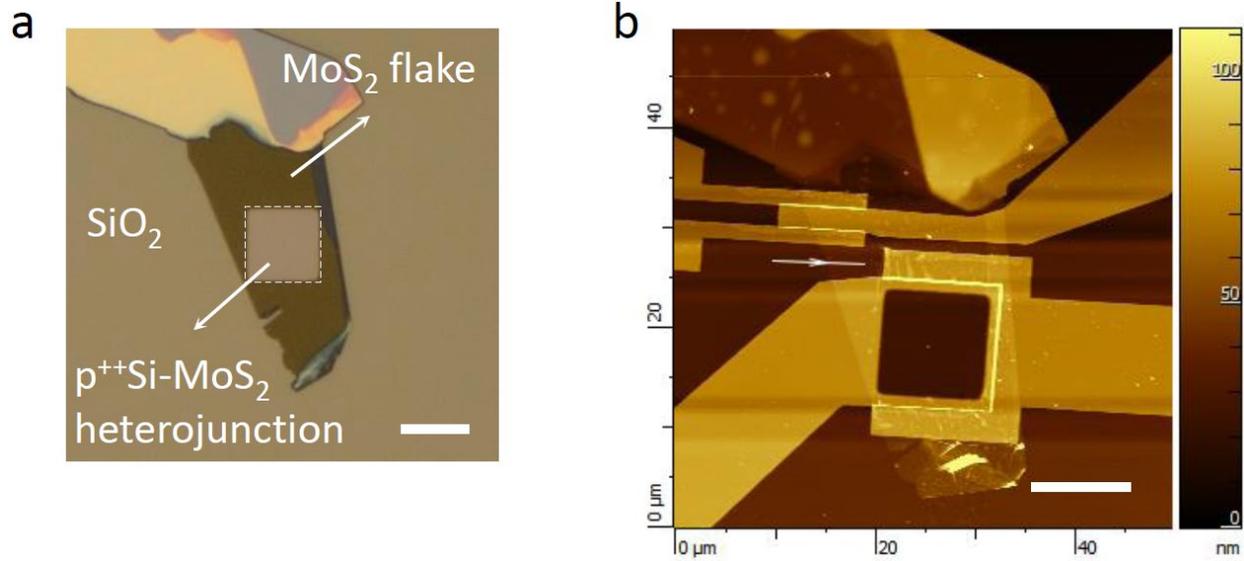

**Figure S2. Optical micrograph and AFM topography image of a p$^{++}$Si-MoS$_2$ heterojunction diode.** (a) Few-layer MoS$_2$ flake covering a square groove where inside groove is p$^{++}$Si and outside is 50-nm thick SiO$_2$. Dashed square outlines the p$^{++}$Si-MoS$_2$ heterojunction. Scale bar, 10 μm. (b) AFM topography image of a p$^{++}$Si-MoS$_2$ diode built from 2D-3D heterostructure in Fig. S2a indicates clean 2D-3D junction interface. Scale bar, 10 μm.

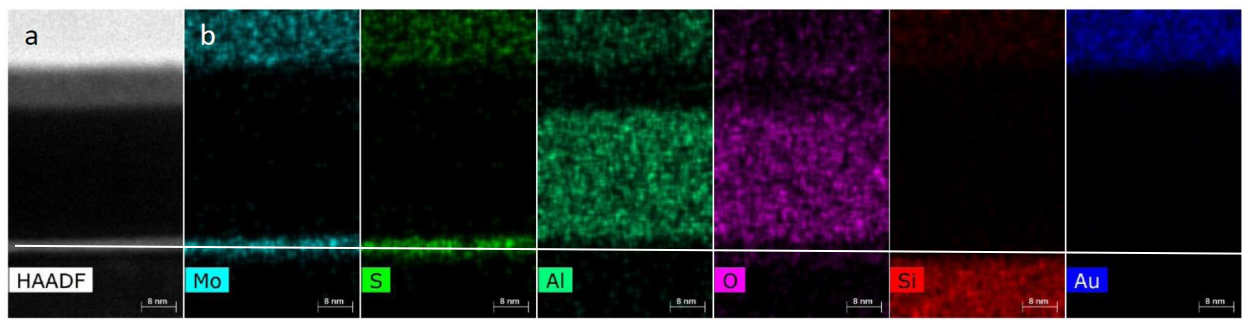

**Figure S3. TEM characterization of a p$^{++}$Si-MoS$_2$ heterojunction diode**. (a) Cross-sectional TEM image of a representative p$^{++}$Si-MoS$_2$ heterojunction diode showing Si, MoS$_2$, Al$_2$O$_3$ gate dielectric, and Ti/Au gate electrode. Scale bar, 8nm. (b) Electron energy-loss spectroscopy (EELS) map of the heterojunction region showing spatial distribution of molybdenum, sulfur, silicon, aluminum elements, thus confirming the location of the Si, MoS$_2$, Al$_2$O$_3$ in the device. Scale bar, 8 nm. TEM sample was prepared by focused ion beam with 5 keV Ga$^+$ ions for final milling. For EDX mapping, a Talos F200X equipped with a four-quadrant 0.9-sr energy-dispersive X-ray spectrometer operated at 200 keV was used.

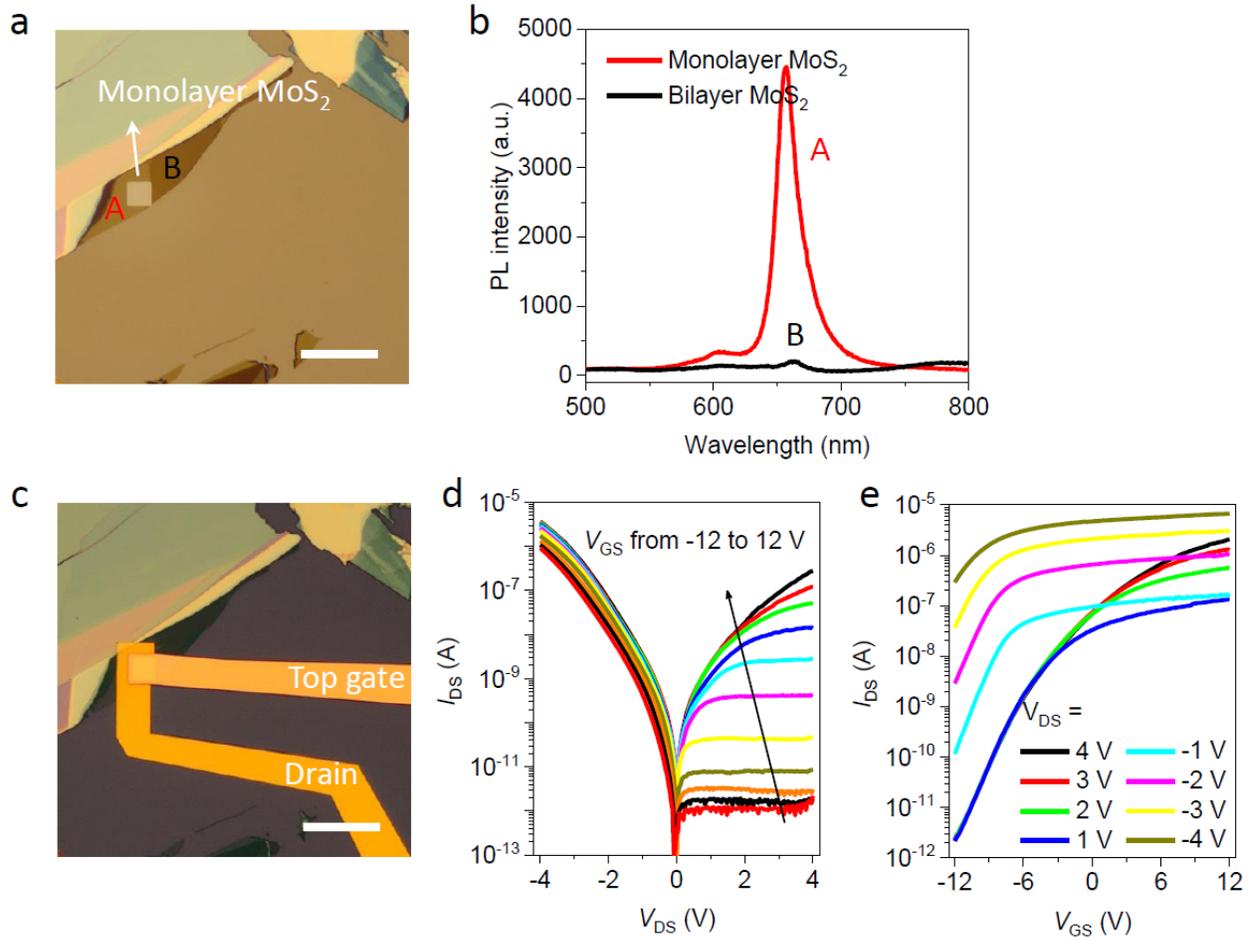

**Figure S4. Electrical characterization of monolayer MoS$_2$-p$^{++}$Si 2D/3D junction diode.** (a) Optical micrograph of monolayer MoS$_2$-p$^{++}$Si heterostructure. scale bar, 20 μm. (b) Photoluminescence spectrum of monolayer and bilayer MoS$_2$ taken from region A and B of the panel a. Monolayer MoS$_2$ is a direct bandgap semiconductor showing strong photoluminescence (red). Bilayer MoS$_2$ is an indirect bandgap semiconductor showing very weak photoluminescence (black). (c) Optical micrograph image of monolayer MoS$_2$-p$^{++}$Si heterojunction device. scale bar, 20 μm. (d) Output characteristics of the device at various gate voltages indicating a rectification ratio of 10$^6$. (e) Transfer characteristics of the device at various drain biases showing a maximum on/off ratio of 10$^6$.

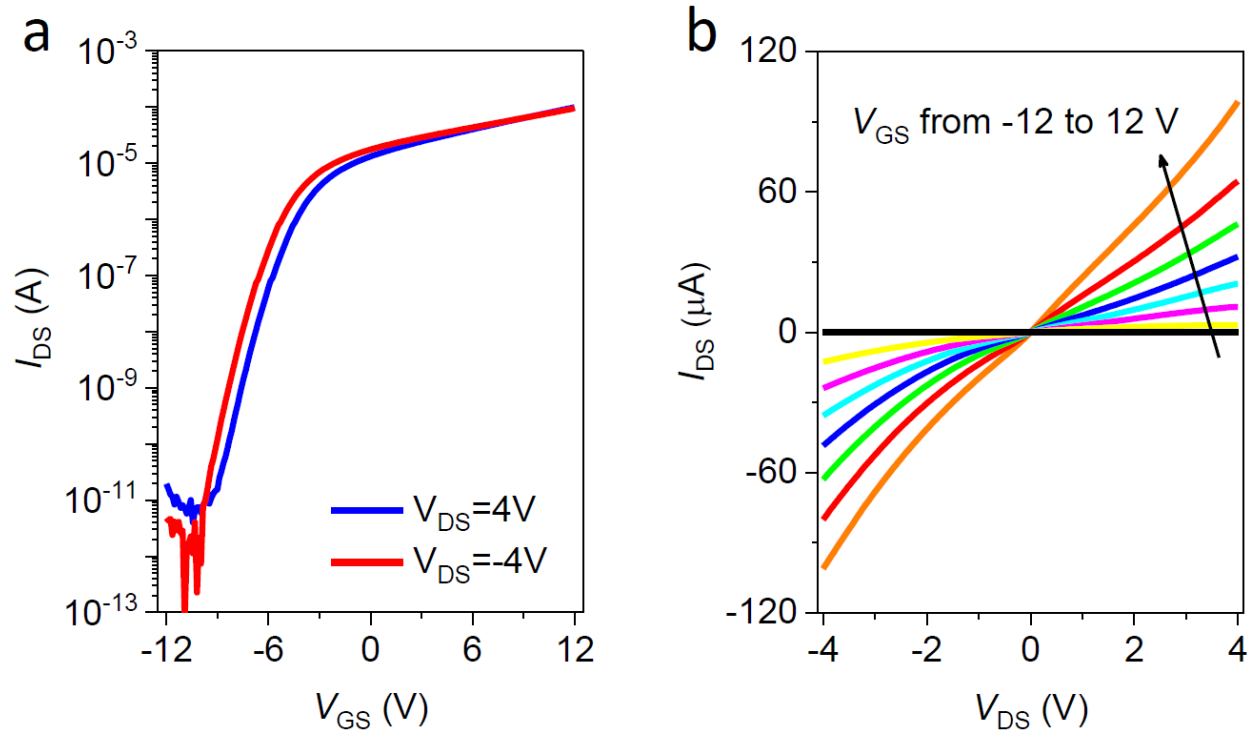

**Figure S5. Room-temperature electrical characterization of a MoS$_2$ FET.** The MoS$_2$ FET was made from the same fabrication process as p$^{++}$Si-MoS$_2$ heterojunction diodes. Source, drain and gate electrodes are made from e-beam deposited Ti/Au (10nm/40nm). A 30-nm thick Al$_2$O$_3$ gate dielectric is deposited by ALD process. (a) Transfer characteristics of a MoS$_2$ FET at $V_{DS}$ = 4V (blue) and $V_{DS}$ = -4V (red). The device on/off ratio is 10$^7$ and mobility is ~ 5.2 cm$^2$/Vs. (b) Output characteristics for various gate voltages changing from -12 to 12V, with a step size of 3V. Output characteristics of the MoS$_2$ FET show ohmic-like I-V behavior indicating that the rectifying I-V behavior of p$^{++}$Si-MoS$_2$ heterojunction diode comes from the 2D-3D heterojunction instead of metal/MoS$_2$ junction.

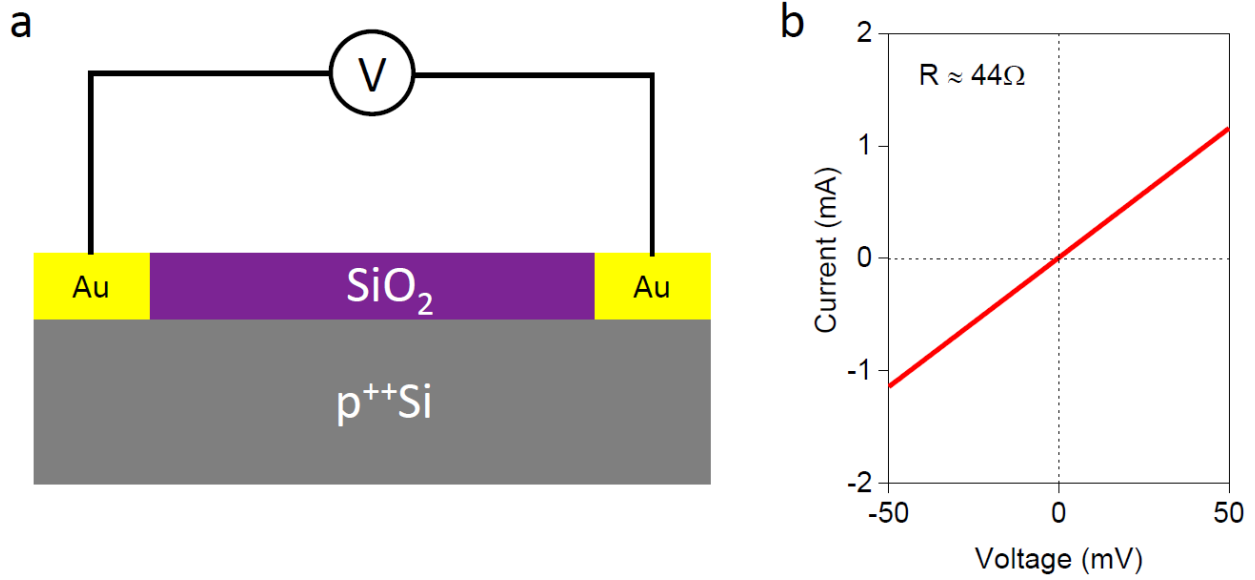

**Figure S6. Electrical characterization of Au/p$^{++}$Si/Au two-terminal device**. (a) Schematic of Au/p$^{++}$Si/Au two-terminal device. (b) Current versus voltage characteristics showing extremely small resistance of ~ 44 Ω.

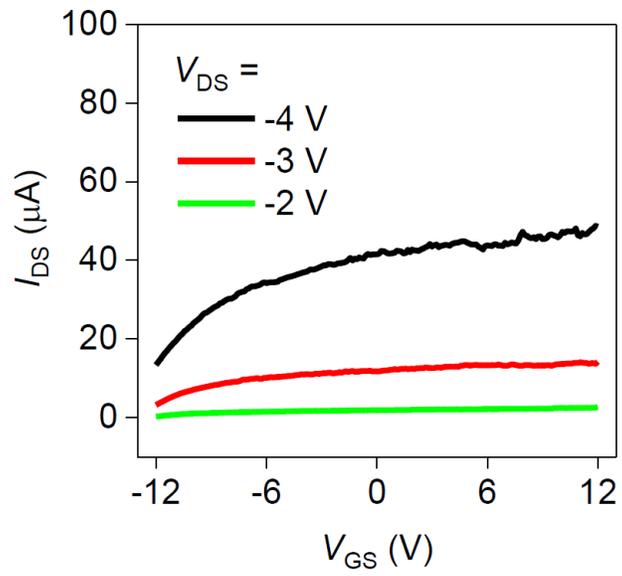

**Figure S7. Linear scale transfer characteristics of a p⁺⁺Si-MoS$_2$ heterojunction diode**. Under forward drain bias ($V_{DS}$ = -4, -3, -2V), there is little or no current modulation as a function of gate voltage.

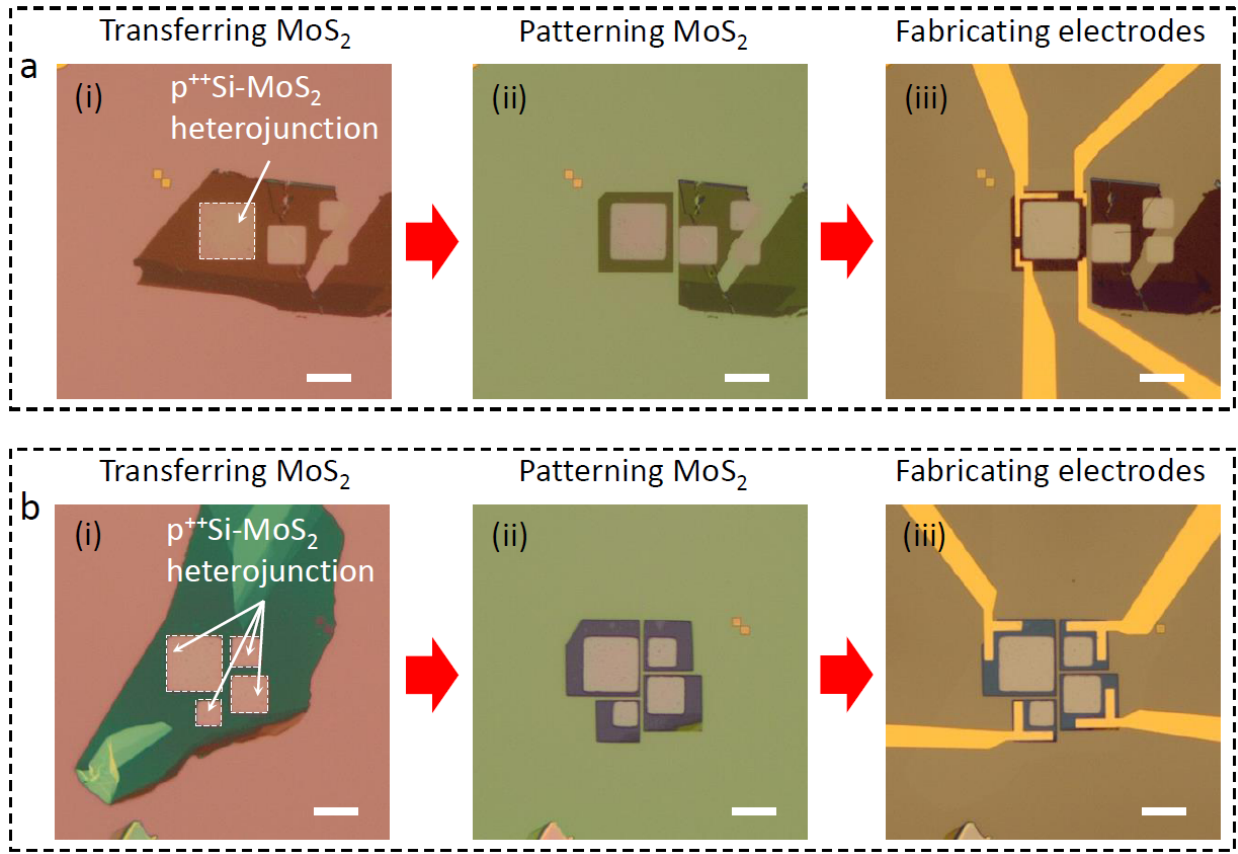

**Figure S8. Fabrication of p⁺⁺Si-MoS₂ heterojunction diodes with different contact lengths and junction areas.** (a) A single MoS$_2$ flake transferred onto p$^{++}$Si square groove (i), patterned by XeF$_2$ gas (ii) and fabricated four metal electrodes onto patterned MoS$_2$ Flake. Scale bar, 10 μm. (b) A single MoS$_2$ flake transferred onto four p$^{++}$Si square grooves (i), patterned by XeF$_2$ gas (ii) and fabricated four metal electrodes onto MoS$_2$ flakes. Scale bar, 10 μm. Unwanted MoS$_2$ regions were etched away by XeF$_2$ etcher (XeF$_2$ pressure: 3.0 Torr, N$_2$ pressure: 2.0 Torr, Etch time: 15 sec/cycle, Number of cycles: 6).

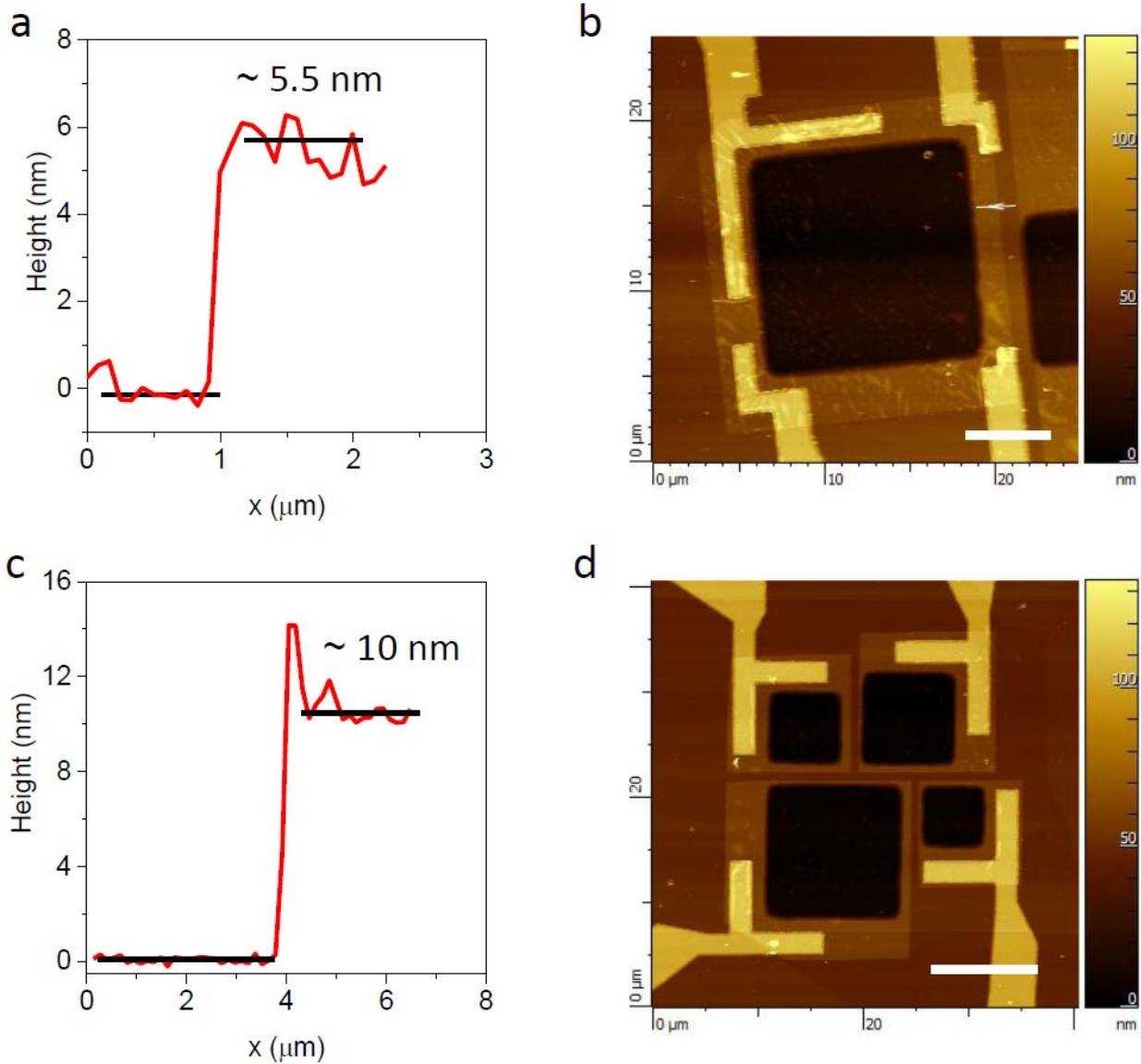

**Figure S9. AFM characterizations of p$^{++}$Si-MoS$_2$ heterojunction diodes.** (a) Height profile across MoS$_2$ flake indicating that the flake is ~ 5.5-nm thick. (b) AFM topography image of the p$^{++}$Si-MoS$_2$ heterojunction diode with different metal contact lengths built from a single MoS$_2$ flake. It demonstrates very clean and uniform van der Waals contacts. Scale bar, 5 μm. (c) Height profile across MoS$_2$ flake indicating that the flake is ~ 10-nm thick. (d) AFM topography image of four p$^{++}$Si-MoS$_2$ heterojunction diodes with same metal contact length built from a single MoS$_2$ flake. It demonstrates very clean and uniform van der Waals contacts. Scale bar, 10 μm.

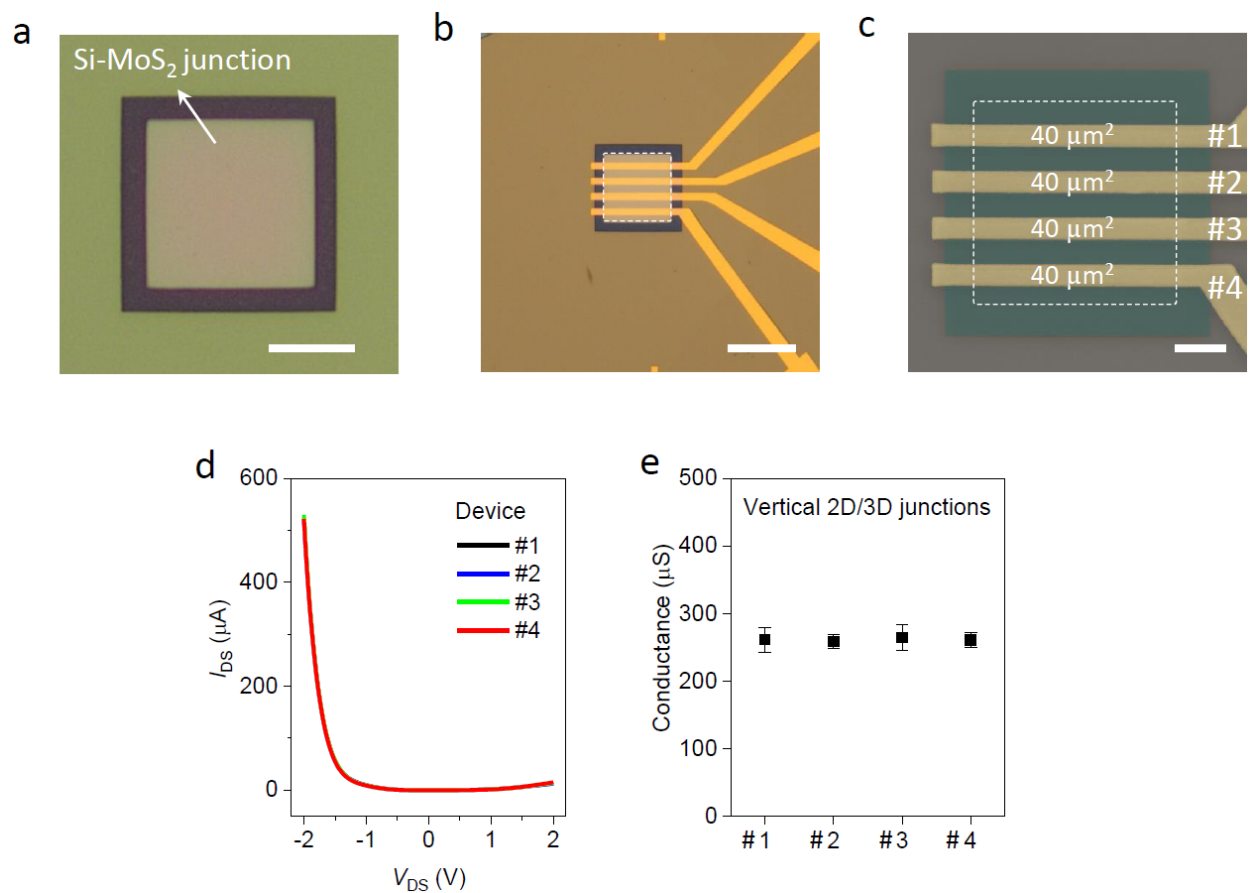

**Figure S10. Electrical characterization of uniform area vertical p$^{++}$Si-MoS$_2$ 2D/3D junction diode.** (a) Optical micrograph of p$^{++}$Si-MoS$_2$ heterostructure showing a very clean interface. Scale bar, 10 μm. Optical micrograph (b) and SEM image (c) of a vertical p$^{++}$Si-MoS$_2$ heterojunction device with four top drain electrodes (marked as 1, 2, 3, and 4). These electrodes form the same overlapping junction area and the area is 40 μm$^2$. Scale bar, 20 μm and 5 μm. (d) Output characteristics of four vertical p$^{++}$Si-MoS$_2$ heterojunction devices showing almost same I-V behaviors indicating uniform interface across the 2D/3D junction. (e) Device conductance extracted from panel d showing almost the same values (~260 μS) of the four vertical 2D/3D devices which further demonstrates the uniform contacts between MoS$_2$ and p$^{++}$Si.

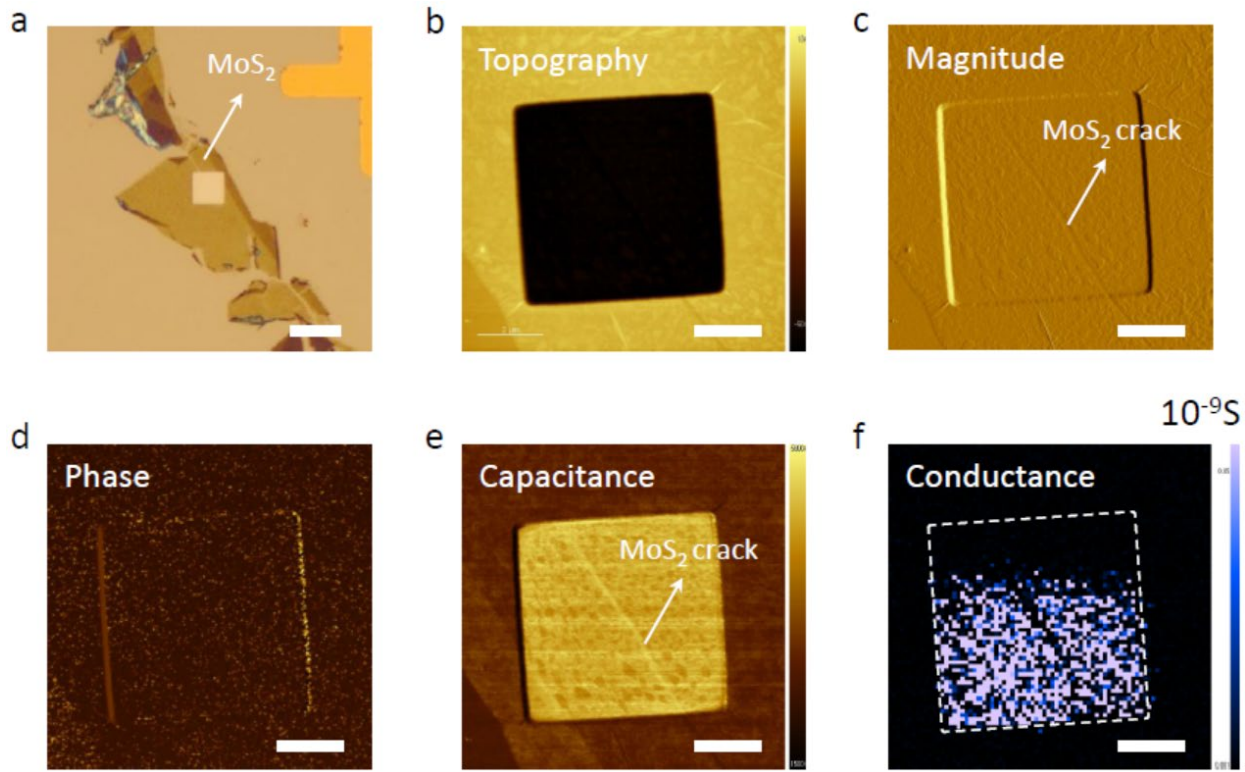

**Figure S11. AFM characterization of p$^{++}$Si-MoS$_2$ heterostructure**. (a) Optical micrograph of p$^{++}$Si-MoS$_2$ heterostructure. Scale bar, 10 µm. AFM topography (b), magnitude (c) and phase (d) image of p$^{++}$Si-MoS$_2$ heterostructure showing very clean 2D/3D interface. Scale bar, 2 µm. (e) AFM capacitance characterization of p$^{++}$Si-MoS$_2$ heterostructure showing very uniform capacitance across the 2D/3D junction. Scale bar, 2 µm. (f) Conductive AFM map of p$^{++}$Si-MoS$_2$ heterostructure. Scale bar, 2 µm.

We have used conductive AFM to determine the 2D/3D junction uniformity and vertical current transport as shown above in **Figure S11**. AFM topography, magnitude and phase characterizations show good interface uniformity across the 2D/3D junction region as shown in **Figure S11b-d**. The uniform AFM capacitance map (**Fig. S11e**) also indicates a good AC electrical interface between MoS$_2$ and p$^{++}$Si. The DC conductance map **Figure S11f** also shows uniformly high conductance in the square that forms the junction except for places where MoS$_2$ has formed a winkle or crack or in places where topography prevents the tip from making a stable electrical contact such as the edges of the pattern. We observe the uniform conductance across the 2D/3D junction which is outlined by the white dashed line. The measured conductance is point by point conductance measured in a purely vertical geometry where the tip is biased and sample (p$^{++}$ Si is grounded). However, the conductive AFM tip is sensitive to oxidation and wear for prolonged measurements under ambient conditions. Therefore, unfortunately the conductive AFM tip wears out towards the end of the scan (scan performed from bottom to top) and hence the top part of conductance is missing as seen in **Figure S11f**.

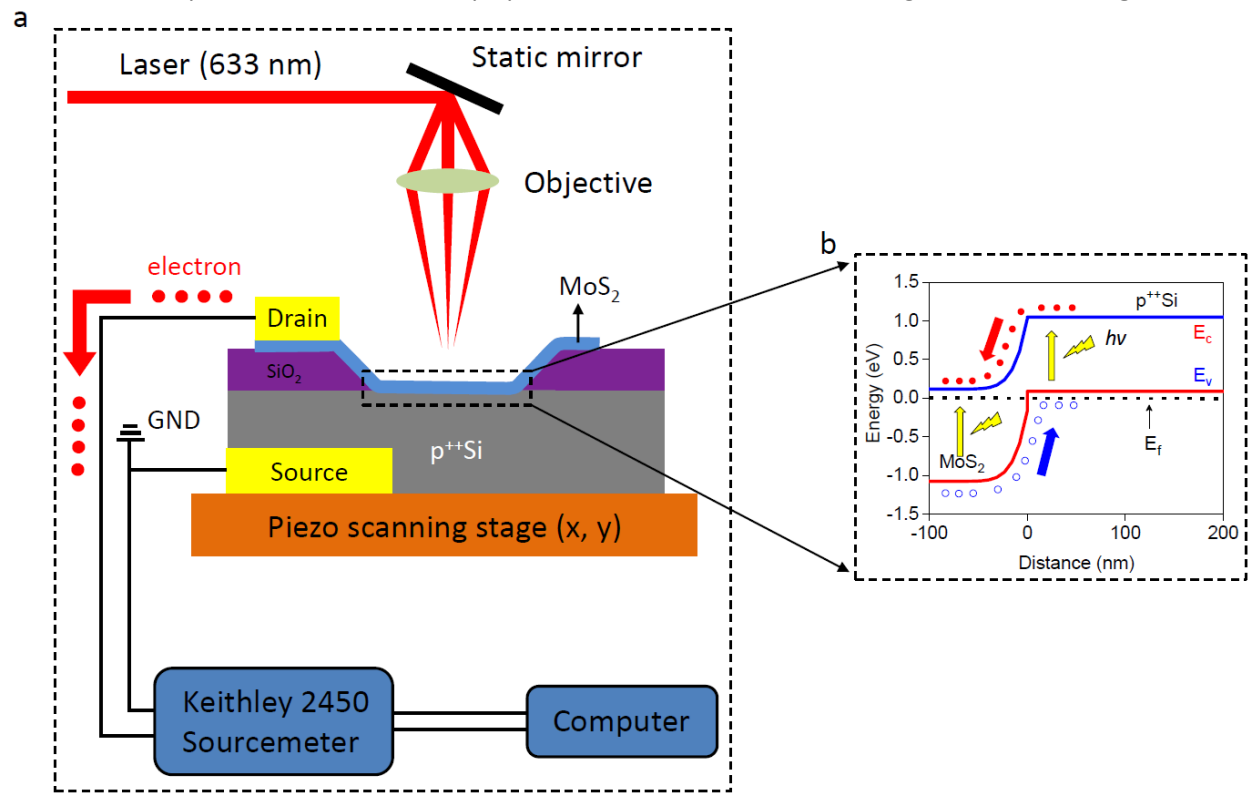

**Figure S12. Scanning photocurrent setup and 2D/3D band diagram under illumination.** (a) Scanning photocurrent setup with focused laser spot (several μm$^2$) on the 2D/3D device. The device is attached onto piezo scanning stage (orange box). The Keithley 2450 Sourcemeter is used to collect photo-generated carriers where MoS$_2$ is the Drain and p$^{++}$Si is grounded (Source). Therefore, the sourcemeter collects photo-generated electrons (red balls) from the device leading to negative photocurrent. (b) Device band diagram under laser illumination. The photo-generated electrons (red balls) flow to MoS$_2$ side and photo-generated holes (blue rings) flow to p$^{++}$Si side due to the potential difference.

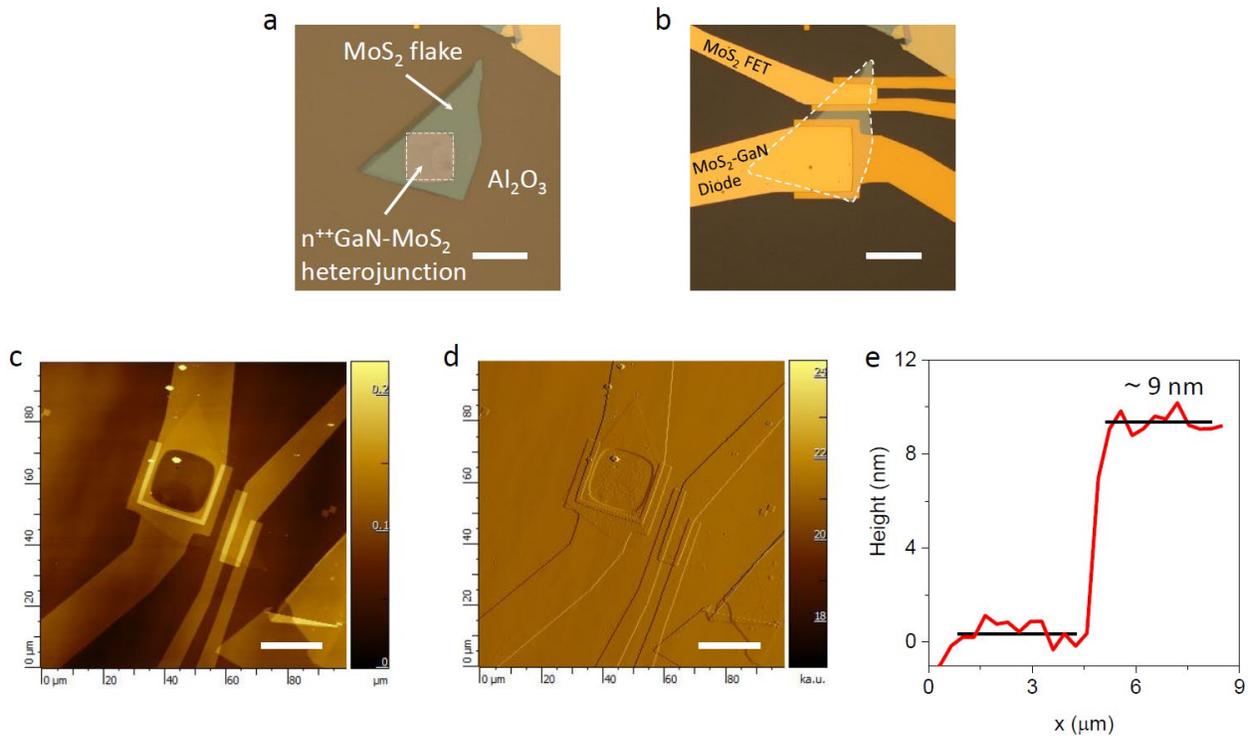

**Figure S13. Optical micrograph and AFM images of n$^{++}$GaN-MoS$_2$ heterojunction diodes.** (a) Optical micrograph of a MoS$_2$ flake covering a square groove where inside groove is n$^{++}$GaN and outside is 60-nm thick Al$_2$O$_3$ oxide layer. Dashed square outlines the n$^{++}$GaN-MoS$_2$ heterojunction. Scale bar, 20 μm. (b) Optical micrograph of a representative n$^{++}$GaN-MoS$_2$ heterojunction diode built from the heterostructure in Fig. S13a. Dashed line outlines a MoS$_2$ flake. Two types of devices, such as a MoS$_2$ FET and an n$^{++}$GaN-MoS$_2$ heterojunction diode, are built from the same MoS$_2$ flake. Scale bar, 20 μm. (c) AFM topography image of the n$^{++}$GaN-MoS$_2$ heterojunction diode. Scale bar, 20 μm. (d) AFM magnitude image of the n$^{++}$GaN-MoS$_2$ heterojunction diode. Scale bar, 20 μm. (e) AFM height profile across MoS$_2$ flake of the n$^{++}$GaN-MoS$_2$ heterojunction diode. The MoS$_2$ flake is ~ 9-nm thick.

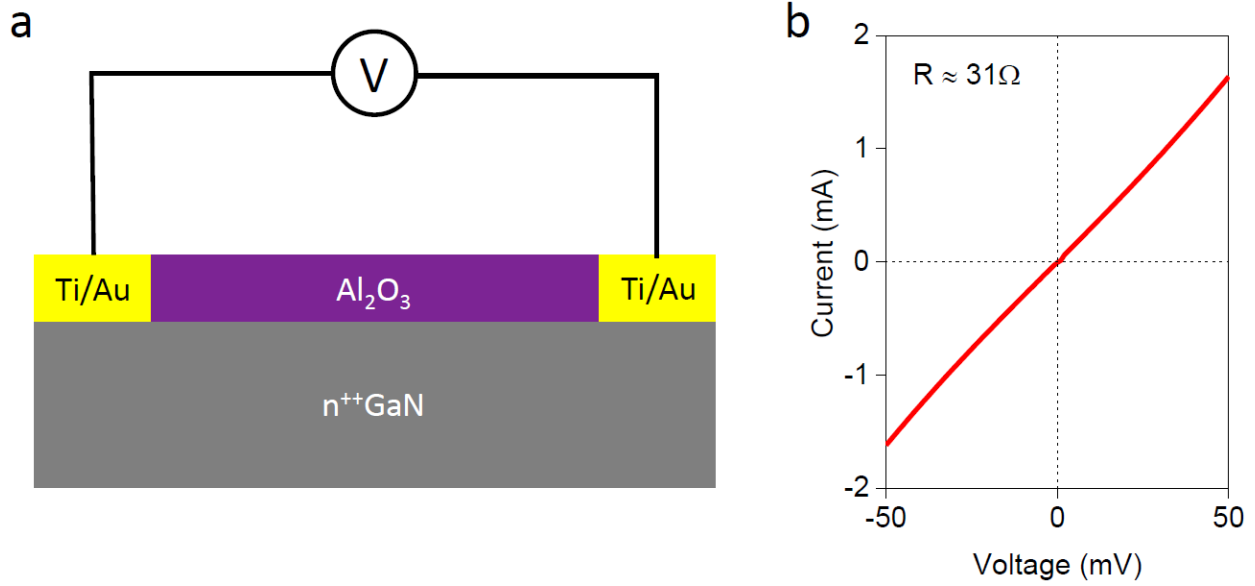

**Figure S14. Electrical characterization of Ti/Au-n$^{++}$GaN-Ti/Au two-terminal device.** (a) Schematic of Ti/Au-n$^{++}$GaN-Ti/Au two-terminal device. (b) Current versus voltage characteristics showing extremely small resistance of ~ 31 Ω.

## Section S1. Ideality factors of 2D/3D heterojunction diodes

Here, we extract the ideality factors of 2D-3D heterojunction diodes using the logarithmic of the forward-biased I-V curves based on the following Shockley diode equation:

$$I = I_{sat}\left(\exp\left(\frac{qV}{nk_BT}\right) - 1\right)$$

Where I is the diode output current, V is the drain bias, $I_{sat}$ is the saturation current, q is the elementary charge, T is the temperature in Kelvin, $k_B$ is the Boltzmann constant and n is the ideality factor. The -1 term is negligible for $V_{DS}$ > 0.1V. And the above equation is reduced to

$$I = I_{sat}\left(\exp\left(\frac{qV}{nk_BT}\right)\right)$$

If we take the log of both sides of the above equation, it gives the following equation

$$\ln(I) = \ln(I_{sat}) + \frac{qV}{nk_BT}$$

As can be seen, ln (I) linearly depends on drain voltage and the slope is $q/nk_BT$. Figure S16b shows the ideality factor of a p$^{++}$Si-MoS$_2$ heterojunction diode as functions of drain bias and gate voltage. The smallest ideality factor is around 2.0 at $V_{DS}$ > -0.1V and $V_{GS}$ = -12V. Furthermore, the smallest ideality factor of a n$^{++}$GaN-MoS$_2$ heterojunction diode is around 1.6 at $V_{DS}$ > -0.1V and $V_{GS}$ = -12V (Fig. S16d).

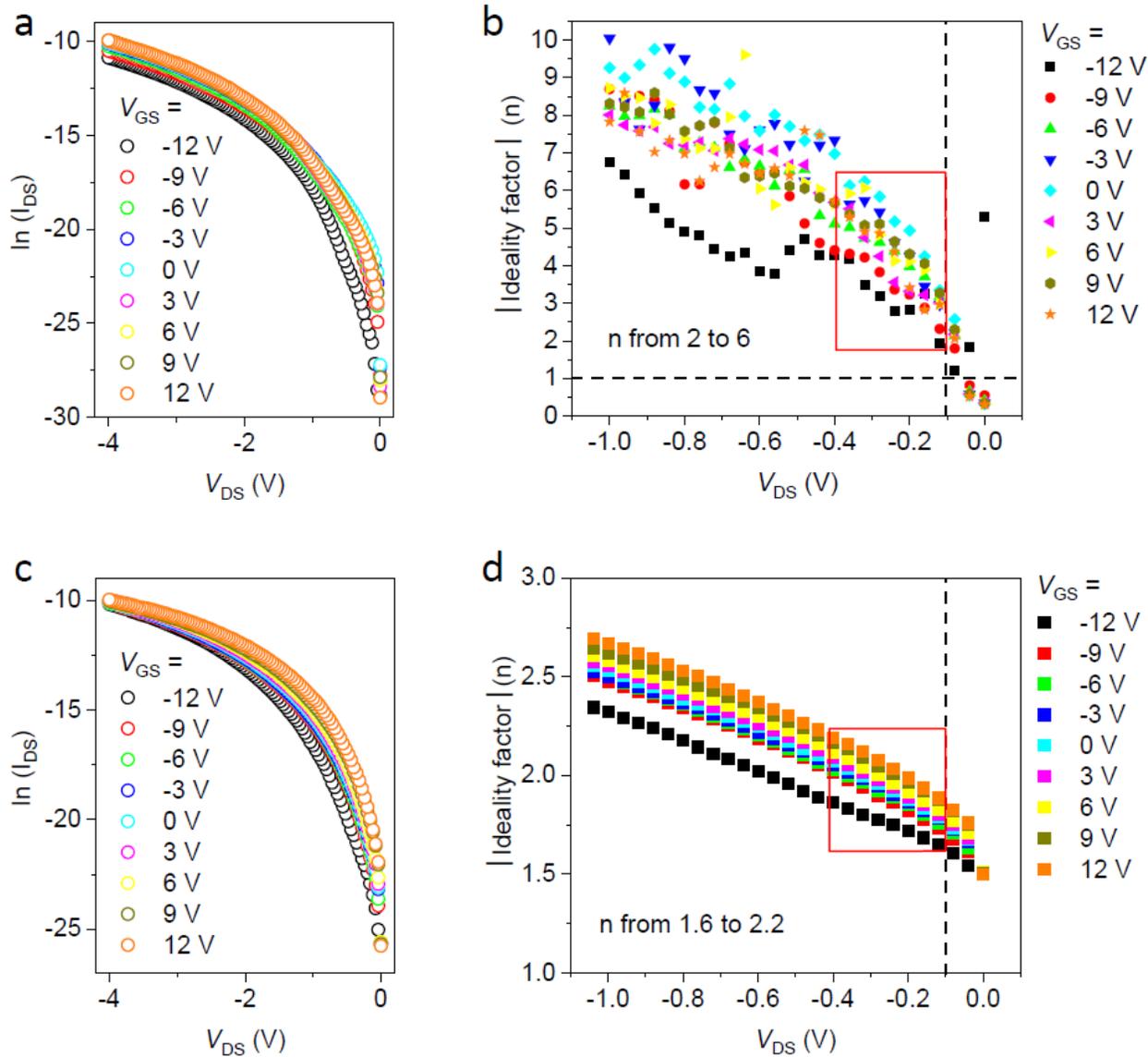

**Figure S15. Ideality factors of p$^{++}$Si-MoS$_2$ and n$^{++}$GaN-MoS$_2$ heterojunction diodes at room temperature.** (a) ln (I$_{DS}$) versus V$_{DS}$ as a function of gate voltage. The gate voltages change from -12 to 12 V, with a step size of 3V. (b) Ideality factors of a p$^{++}$Si-MoS$_2$ heterojunction diode as functions of drain bias and gate voltage. (c) ln (I$_{DS}$) versus V$_{DS}$ as a function of gate voltage. The gate voltages change from -12 to 12 V, with a step size of 3V. (d) Ideality factors of a n$^{++}$GaN-MoS$_2$ heterojunction diode as functions of drain bias and gate voltage.